\documentclass[11pt,a4paper]{quantumarticle}
\pdfoutput=1

\usepackage{amsmath}
\usepackage{amssymb}
\usepackage{booktabs}
\usepackage{graphicx}
\usepackage[english]{babel}
\usepackage{siunitx}
\usepackage{xcolor}
\usepackage{hyperref}
\usepackage{placeins}
\usepackage{orcidlink}

\newcommand\beq{\begin{equation}}
\newcommand\eeq{\end{equation}}

\definecolor{comment}{gray}{0.4}

\begin{document}

\title{Training Neural Networks with Universal Adiabatic Quantum Computing}

\begin{flushright}
    {IPPP/23/46, CERN-TH-2023-162}
\end{flushright}

\author{Steve Abel \orcidlink{0000-0003-1213-907X}}
\email{s.a.abel@durham.ac.uk}
\affiliation{Institute for Particle Physics Phenomenology, Durham University, Durham DH1 3LE, UK}
\affiliation{Theoretical Physics Department, CERN,
Esplanade des Particules 1, Geneva CH-1211, Switzerland}

\author{Juan Carlos Criado \orcidlink{0000-0003-3571-994X}}
\email{juan.c.criado@durham.ac.uk}
\affiliation{Departamento de F\'isica Te\'orica y del Cosmos,
Universidad de Granada, E–18071 Granada, Spain}

\author{Michael Spannowsky \orcidlink{0000-0002-8362-0576}}
\affiliation{Institute for Particle Physics Phenomenology, Durham University, Durham DH1 3LE, UK}
\email{michael.spannowsky@durham.ac.uk}

\maketitle

\begin{abstract}
The training of neural networks (NNs) is a computationally intensive task requiring significant time and resources. This paper presents a novel approach to NN training using Adiabatic Quantum Computing (AQC), a paradigm that leverages the principles of adiabatic evolution to solve optimization problems. We propose a universal AQC method that can be implemented on gate quantum computers, allowing for a broad range of Hamiltonians and thus enabling the training of expressive neural networks. We apply this approach to various neural networks with continuous, discrete, and binary weights. Our results indicate that AQC can very efficiently find the global minimum of the loss function, offering a promising alternative to classical training methods. 

\end{abstract}

\flushbottom

\section{Introduction}
\label{intro}

Adiabatic quantum computing (AQC) is a paradigm of quantum computation that harnesses the principle of adiabatic evolution to solve computational problems \cite{farhi2000, farhi01a, farhi02a}. In this approach, the quantum system is initialized in the ground state of a simple Hamiltonian. The system is then evolved adiabatically, ensuring that it remains in its ground state towards a final Hamiltonian which encodes the solution to the problem. The adiabatic theorem guarantees that if the evolution is sufficiently slow, the system will remain in the ground state throughout the process. The computational prowess of AQC is equivalent to that of the conventional quantum computation model, implying that both models are polynomially equivalent \cite{aqcequiv}: in other words it is considered to be a universal quantum computing paradigm. Moreover, AQC has been realized experimentally in various systems, including solid-state single-spin systems under ambient conditions \cite{aqcexp,Biamonte2008}. The purpose of this paper is to demonstrate how AQC can be used to greatly enhance Neural Networks.

Generally, NNs, like all self-adaptive optimisation algorithms, consist of three core parts:
\begin{enumerate}
\item A system that encodes a complex function,
\item An output layer's loss function that dictates the NN's task,
\item A training method to minimize the loss function.
\end{enumerate}
It is the last of these three, namely the training of NNs, which typically demands the greatest time, effort and resource, and which poses the greatest challenge to their development and deployment.

In previous exploratory work \cite{Abel:2022lqr,Abel:2020qzm,Abel:2020ebj} we showed that a NN can be trained  by encoding it in a transverse Ising model on a quantum annealer \cite{LantingAQC2017}. That 
work demonstrated that such an approach, utilising quantum tunnelling, can train a NN optimally, reliably and quickly. Furthermore, the trained parameters can be extracted and used in a classical network for deployment. However, the restriction to a transverse Ising model as the Hamiltonian for quantum annealing greatly limits the expressivity of the NN.

Thus, to address these obstacles, this paper proposes a universal AQC approach that can be used to train a neural network and extract the optimally trained network weights. The much wider variety of Hamiltonians that can be used within the universal AQC paradigm allows us to include correlations and non-linearities in our models, allowing adiabatic quantum training to be applied to  larger and more expressive networks. 

We will present two techniques for performing the AQC, the ``matrix method'' in which the system is expressed in terms of truncated Hilbert space components, and the ``Pauli-spin method'' in which it is expressed directly with Pauli-spin matrices. 
We apply these methods to simulated quantum-gate computers, showing the applicability of AQC training on near-term devices. Furthermore, we apply the ``Pauli-spin method'' to the training of both continuous neural networks, and networks with discrete and binary weights. The latter usually rely on non-gradient-based optimisation algorithms and are classically very difficult to treat.

In the burgeoning domain of computational intelligence, neural networks are heralded as the cornerstone of machine learning, particularly excelling in classification and regression tasks. Their influence permeates both everyday applications and advanced scientific research. Hence, being able to enhance their capabilities and streamline their training through the innovative lens of quantum computing is of considerable significance.

\section{Challenges in Training Neural Networks}
\label{sec:nntrain}

It is useful to begin our discussion  with a brief appraisal of the difficulties one may encounter when training a neural network. In the training phase the goal is to reach the global minimum of the so-called cost or loss function. However, the optimization landscape of neural networks often contains multiple local minima. This problem is only exacerbated if the network is a small one. Broadly speaking, in a space of high dimensionality most critical points are likely to be saddle points. Thus a gradient descent method is usually effective. Conversely on small neural networks finding the global minimum can be much more difficult. Indeed several other issues can arise during training, even with correct algorithm implementation. Here, we briefly list these challenges and the typical approach that is employed to deal with them in classical training:

\begin{itemize}

\item{Slow Progress, Fluctuations or Instability}: tackled by optimising the learning rate to either speed up slow down convergence.

\item{Badly Conditioned Curvature}: ``ravines'' in the landscape imply that different directions need different learning rates to be optimal. The Adam algorithm can address this by adapting learning rates individually for each parameter.

\item{Local Optima}: addressed by using  random restarting points to explore every basin of attraction. 

\item{Weight degeneracy}: addressed by a random initialization of the weights and biases, which breaks the symmetry. 

\item{Dead and Saturated Units}: activations at the ends of their range cause plateaus in the loss-function landscape. Initializing biases with positive values can help avoid the problem, although it can also signal a redundancy in the network that one would like to reduce by pruning out redundant weights. 

\end{itemize}

It is worth emphasising that the paradigm of neural networks uses a set of {\it continuous} weights and biases on which a gradient descent can be performed. However, arguably this causes great redundancy because, in many situations, a reasonable solution to the optimisation of the network is, in principle, achievable with weights that are discrete or even binary (i.e. just ``on'' or ``off'') if only we can find the correct discrete values.

To appreciate the redundancy that is inherent in continuous weights and biases consider the example of a classification task when there are only two features. In principle, the classification curve can be written as the Taylor expansion of the level-curve of some function $z(x_1,x_2)$ of the features $x_1$ and $x_2$. However, if this classification curve happens to be well approximated by a quadratic function for example, then it would require only six continuous coefficients. In contrast, the neural network would typically have many more continuous weights and biases. Conversely, if we accept that these six continuous Taylor coefficients are well approximated if we know them to four binary places (i.e. to one part in 16) then only 24 binary weights taking values of 0 or 1 should {\it in principle} be able to describe the same classification curve. Because of this redundancy, there is indeed quite some interest in training discretely weighted networks, and networks where both the weights and activation functions are binary \cite{2021arXiv211006804Y,roth,binaryNN,2015arXiv151100363C,2019arXiv190402823D}.

However, we can immediately appreciate that such a system of discrete weights is classically problematic precisely because it runs into both the ``weight degeneracy'' and the  ``dead and saturated units'' problems mentioned in our list of challenges. Moreover the ``local optima'' problem is generic. Indeed it is for these reasons that the classical training of discretely weighted and binary systems requires special treatment
\cite{roth,binaryNN}.

\section{Adiabatic Quantum Computing on Gate Quantum Computers }
\label{AQC}

\begin{figure*}
  \centering
  \includegraphics[width=0.49\textwidth]{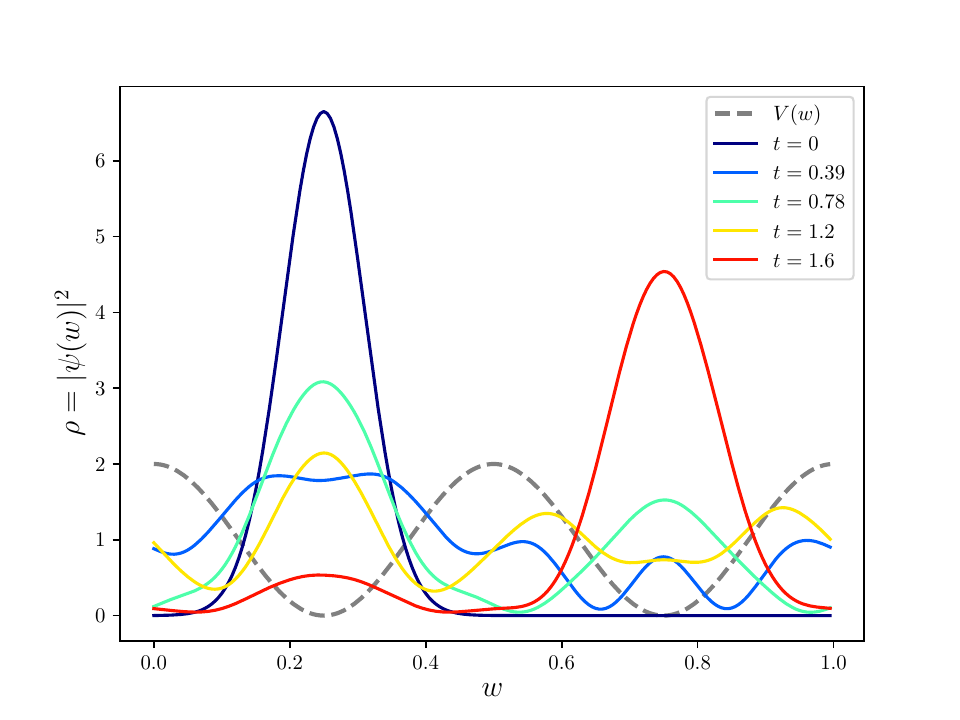}
  \includegraphics[width=0.49\textwidth]{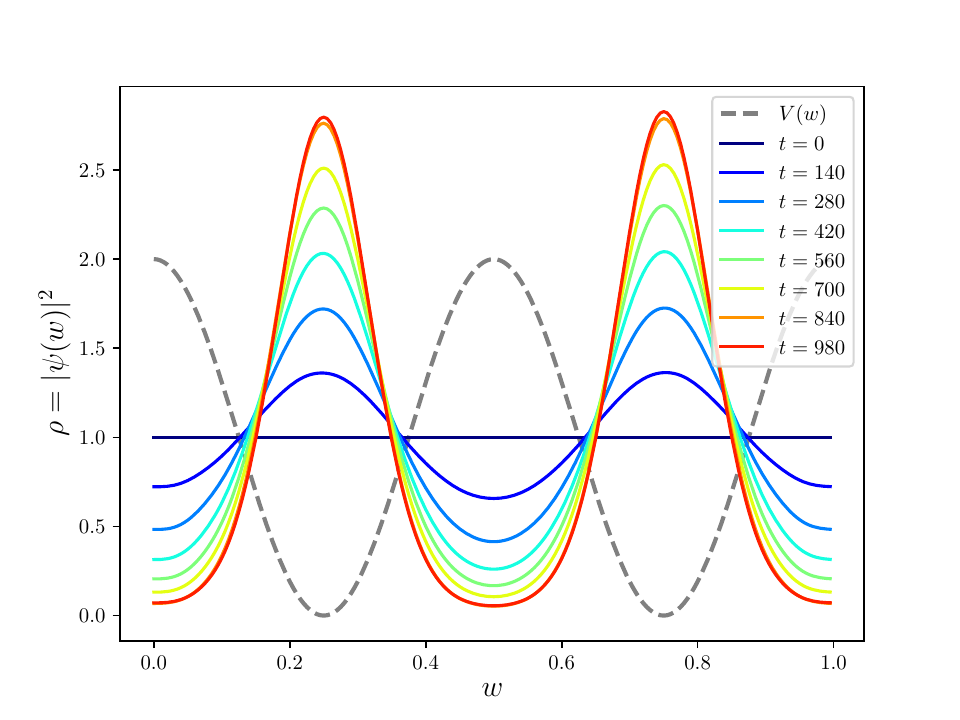}
  \caption{Tunnelling versus adiabatic evolution  in the cosine potential, $V(w)=1+\cos(4\pi w)$, with a truncation at energy level $\langle w|n \rangle   = e^{2 \pi n w i  }$ where $n\in [-15,15]$. In the tunnelling example we take $m=10$, while in the second adiabatic example we have taken masses of $m=100$ (equivalently $V$ can be multiplied by $100$) in order to get well localized peaks in the ground state. The evolution time must be increased accordingly. The  schedule function is taken to be simply linear, $s(t)=t$.} 
  \label{fig:tunnel+anneal}
\end{figure*}

Let us now therefore turn to AQC, beginning with a discussion of its general implementation on gate quantum computers.

Although it is interesting for the reasons outlined above to allow our eventual systems of interest to be relatively discrete in nature, it will be useful in establishing the basic principles first to consider systems of function of continuous variables. Thus in this section, we shall focus on the specific task of finding all the global minima of a function $V(w)$ of one variable in the interval $w\in [0,1]$. (To remind ourselves that ultimately we will be concerned with weights and biases, we call the variable $w$.) AQC is equivalent in this context to solving for the ground state in one-dimensional quantum mechanics with $w$ corresponding to the single space dimension. Studying such familiar cases will allow us to confirm that our system behaves as expected.

The first example we will look at is  the following cosine potential which has two degenerate minima in the interval $w\in [0,1]$:
\beq
V(w)~=~1+\cos (4\pi w)~,
\label{eq:v_first}
\eeq
which appears as the dashed line in Fig.~\ref{fig:tunnel+anneal}. 

 There are various ways in which one might wish to encode the problem of minimising this potential. It is necessary to ensure that the chosen method is both effective and yields an advantage (in the sense that the difficulty does not scale exponentially with the problem size). Here we shall consider two encoding methods, the ``matrix method'' and the ``Pauli-spin method''.

\subsection{The Matrix Method}

The Matrix method is the most direct: it entails evolving the wavefunction from some starting state using the Schr\"odinger Hamiltonian,
\beq
\hat H ~=~ \frac{\hat p ^2}{2m}+  V(\hat w)~~,
\eeq
in its matrix form in a truncated Hilbert space. We proceed as follows. We adopt periodic boundary conditions, and define  a basis of eigenstates of the kinetic piece in the Hamiltonian working in the $w$-basis:
\beq
\langle w | n \rangle ~=~ e^{2\pi i n w}~.
\eeq
The Hamiltonian matrix is then given by 
\begin{align}
H_{n\ell} ~&=~ \int_0^1 \langle n|w\rangle  \langle w|\hat H|\ell \rangle dw\nonumber \\
~&=~  \frac{4\pi^2 n^2}{2m} + \widetilde V(n-\ell) ~~,
\end{align}
where $\widetilde V (n)= \int_0^1 V(w) e^{-2\pi i n w} dw$
is the Fourier transform of $V(w)$ which we can easily calculate for any $n,\ell$.
Thus we can in principle simply take the resulting matrix $H_{n\ell}$,  and use it to evolve the wavefunction $\psi(w,t)$ from an initial state $\psi(w,0) =   c_n(0)\langle w|n\rangle $, using the Trotterized Sch\"odinger evolution,
\begin{align}
c_n(t)~&=~e^{-iH_{n\ell} t}c_\ell(0)\nonumber \\
~&\approx ~ \left(e^{- i H_{n\ell} \delta t}\right)^{t/\delta t} c_\ell(0)
~.
\end{align}

Up to this point, everything is simple quantum mechanics. However, we wish to encode the wavefunction and its evolution in terms of qubits. This can be done by truncating the Hilbert space to size $2^N$ with $n\in [-2^{N-1} ,2^{N-1} ]$. This allows us to identify each index $n$ with one of the $2^N$ possible eigenvalues of $N$ tensored qubits. The simplest choice for this identification is to treat $n$ like the computational-basis index: that is we associate the binary expression for each $n$ with the eigenvalues of the $N$ tensored binary operators 
\beq 
T = \frac{1}{2}({\mathbbm 1}+Z)~,
\label{eq:binary_def}
\eeq
where $Z$ is the Pauli $Z$-spin matrix for each qubit. Thus for example $|n=-2^{N-1} \rangle  \equiv |000\ldots 000\rangle$, $|n=0 \rangle  \equiv |000\ldots 010\rangle $, $|n=3 \rangle  \equiv |110\ldots 010\rangle $ and so forth.

 To perform the time evolution, our $2^N\times 2^N$ Hamiltonian matrix must then be accordingly decomposed into sums of tensor products of the Pauli-spin matrices, which act on the $N$ tensored qubits, and then the evolution of the initial state Trotterized as above.
To implement this step in the process, here and throughout, we will make extensive use of the {\tt qibo} package of programmes which allows fast evaluation of quantum circuits taking full advantage of hardware accelerators \cite{qibo_paper,qibojit_paper,Efthymiou:2023att,Robbiati:2023mbk}. This package allows one to automate the decomposition step and implement the Trotterized time evolution induced by a symbolic Hamiltonian defined in terms of Pauli-spins, which is rendered as a quantum gate circuit. Moreover, simulation is feasible up to an order of 25 qubits. 

As a warm-up exercise, it is interesting to consider an initial wavefunction localised in one of the minima and observe it tunnel to the other degenerate minimum. This is shown for the potential of Eq.~\eqref{eq:v_first} in the first panel in Fig.~\ref{fig:tunnel+anneal}, where for the initial state, we choose a Gaussian localised in the left minimum. We perform the time evolution as a simulation using {\tt qibo}'s ``StateEvolution'' module, which, as we said, produces and evolves the circuit corresponding to the symbolic Hamiltonian. (Importantly {\tt qibo} allows one to put the same Trotter evolution directly onto a real machine.) 

The wavefunction indeed tunnels to the second minimum, as expected. However, in this initial example, we can also see why quantum tunnelling {\it per se} is not always beneficial for locating global minima. There is no energetic dissipation in an idealised setting, so the initial wave function never stops moving unless it is already in an energy eigenstate. It would, for example, be very hard to determine the global minimum if the minima were only slightly non-degenerate. This can be contrasted with dissipative systems such as those utilised in quantum annealers in Refs.~\cite{Abel:2021fpn, Abel:2022lqr,Criado:2022deo,Criado:2022aoo}.

Hence to determine the true global minimum, we can
use AQC as envisaged in Ref.~\cite{farhi00a, farhi01a, farhi02a}. That is, we begin the system in the ground state of a trivial Hamiltonian $\hat H_0$ and adiabatically evolve the system to the complicated Hamiltonian of interest, $\hat H$. As a function of time, the total Hamiltonian $\hat H_A$ for  adiabatic evolution in the AQC paradigm takes the form 
\begin{equation}
\hat H_A(t) ~=~ (1-s(t))\, \hat H_0 + s(t) \, \hat H ~,
\label{eq:HA_1}
\end{equation}
where $s(t)$ is the so-called schedule function with $s(0)=0$ and $s(t_{\rm final}) =1$.
If the evolution is sufficiently adiabatic, the system will always remain in the ground state. The result is the desired ground state of the complicated Hamiltonian of interest. For the present example, we can take $\hat H_0$ to be the purely kinetic Hamiltonian with $V=0$, for which the $n=0$ state, $\langle w|\psi\rangle = \langle w|000\ldots 010\rangle = 1$,  is trivially the groundstate solution. 

We performed the adiabatic evolution within {\tt qibo}  using  
{\small {\tt models.AdiabaticEvolution}, with, for simplicity, the schedule function taken to be linear, $s=t/t_{\rm final}$.
The resulting evolution is shown in the second panel of Fig.~\ref{fig:tunnel+anneal}. Notably, the complicated Hamiltonian's eventual groundstate function is time-independent as it should be and correctly responds to the two minima degenerately. Thus, for locating the global minima, the mass (or, more generally, the kinetic to potential terms ratio in $\hat H$) plays an important role. The higher the mass is relative to $V(w)$, the sharper the peak around the global minima. This is, of course, to be expected because approximating the potential around each minimum, $w_{\rm min}$,  as a simple harmonic oscillator (SHO), the wavefunction is of the form $$\rho_0(w)=|\psi_0|^2 ~\approx~ 
  \left(
  {m}
  \right)^{1/4} e^{- 4 \pi \sqrt{m} (w-w_{\rm min})^2 }~.$$

\begin{figure}
  \centering
\includegraphics[width=0.5\textwidth]{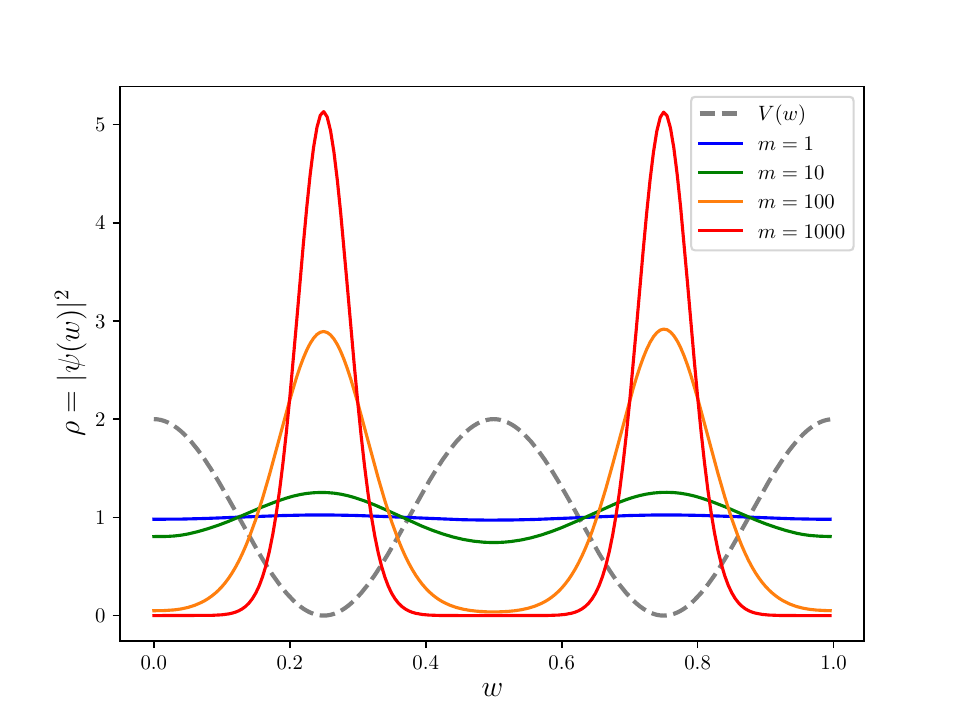}
  \caption{The effect of mass on the ground state. Around each minimum, the groundstate approximates the Gaussian groundstate of the SHO with $V(w) = 8\pi^2 w^2$, namely $\rho_0(w)=|\psi_0|^2 \approx 
  \left(
  {m}
  \right)^{1/4} e^{- 4 \pi \sqrt{m} (w-w_{\rm min})^2 }$ (normalized such that each of the two peaks contributes $1/2$).} 
  \label{fig:groundstate-mass}
\end{figure} 
\noindent We show this dependence explicitly in Fig~\ref{fig:groundstate-mass} which displays the expected $m^{1/4}$ behaviour in the amplitude of the peaks. This feature will be important in later discussions. 

It is instructive and useful for our later discussion to perform the same kind of comparison in a polynomial potential with a metastable minimum where the system can be trapped. A simple case is the following quartic potential:
\begin{equation}
V(w)~=~\lambda(18w^4-35 w^3+22 w^2-5 w+0.372573)~,
\end{equation}
where we keep $\lambda$ as an overall factor to scale the potential. The potential is shown as the grey dashed line in Fig.~\ref{fig:tunnelling}. To examine tunnelling, we begin the system in the approximate ground state of the metastable minimum at $w_+=0.1848$. Expanding around this point, we find an approximate SHO potential with $V(w) \approx \lambda(0.372573+ 2\pi (w-w_+)^2)$ (and hence SHO parameters $m\Omega  = 2\sqrt{\lambda \pi m} $). Thus to demonstrate tunnelling, we begin the system in the Gaussian groundstate,
$$\psi_0(w) ~=~ 
  \left(
  {4m}/{\pi}
  \right)^{1/8} e^{- \sqrt{\lambda \pi m} (w-w_{\rm min})^2 }~.$$
The subsequent evolution is shown in the first panel of Fig.~\ref{fig:tunnelling}. Again we see that tunnelling does not help locate minima without some element of dissipation. Indeed the wavefunction either oscillates wildly between the minima on longer timescales or remains relatively stuck: it is quite challenging to control the behaviour, which depends sensitively on the choice of both $\lambda$ and $m$. This can be contrasted with  AQC which correctly reproduces the ground state in the second panel. This only selects the true global minimum even when the two minima are almost degenerate. As an example of the latter we show in Fig.~\ref{fig:almost_degen} the evolution for the cosine potential (done using the ``matrix method'') with a tiny linear term $\Delta V(w) = \epsilon w$, where $\epsilon=0.02$, which causes non-degeneracy in the two minima. Even though the two minima are imperceptibly non-degenerate, the adiabatic process ultimately finds the true global minimum. The behaviour is quite striking because it is initially degenerate, and only towards the end of the process selects the true minimum. 
\begin{figure}
  \centering
\includegraphics[width=0.5\textwidth]{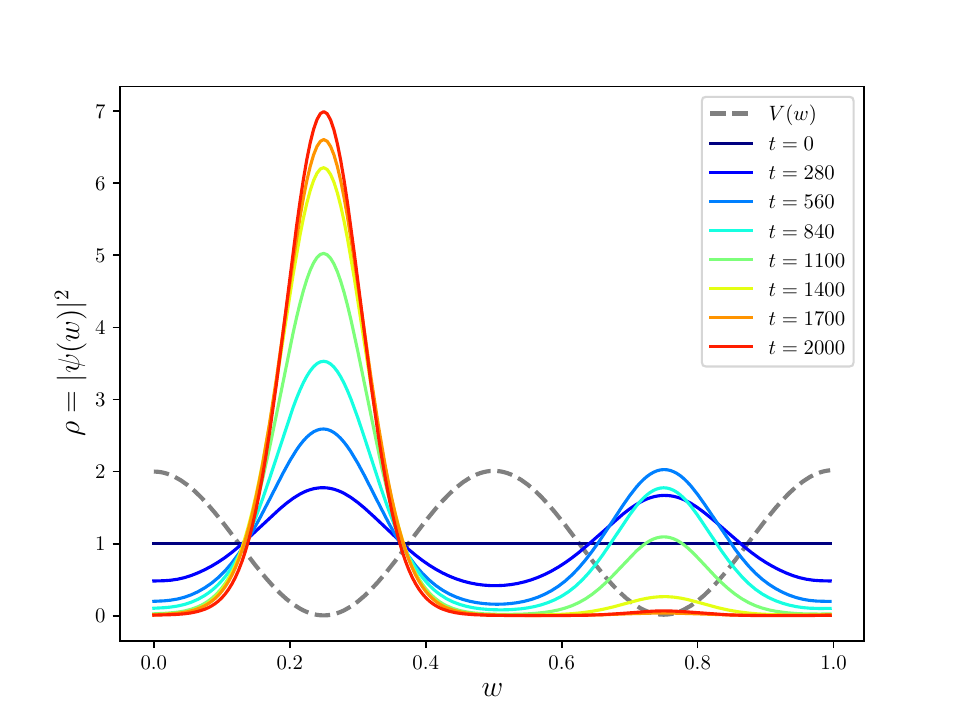}
  \caption{AQC for the exact same system as in Fig~\ref{fig:tunnel+anneal} but with not quite degenerate minima (the energetic difference between the two minima being $\Delta V_{\rm min} \approx 0.01$). During the evolution the ground state ultimately selects the true minimum provided the process remains adiabatic.\label{fig:almost_degen}}
\end{figure} 
As for the cosine potential, the global minimum can be more precisely located by increasing the mass or increasing the parameter $\lambda$, subject to the constraint that the Trotterization should remain a good approximation (i.e. $|H|\delta t\ll 1 $). However in the present context the most important aspect of this example is that we can see that AQC completely avoids the ``Badly Conditioned Curvature'' problem mentioned in our list of challenges in Section~\ref{sec:nntrain}.

We should, for completeness, attach a caveat to this picture: the oscillation back and forth that we can observe in the tunnelling solutions is partly due to the fact that the systems we consider in these illustrative examples are only one-dimensional and periodic. Quantum tunnelling in many physical systems of interest (for example, phase transitions in cosmology) would be higher dimensional and take place in non-compact volumes. In such situations the tunnelling process is one-way because there is a large degenerate volume of global minima: excess energy after tunnelling is dissipated in dynamics, for example, in accelerating bubble walls. 

\begin{figure}[h!]
  \centering
\includegraphics[width=0.5\textwidth]{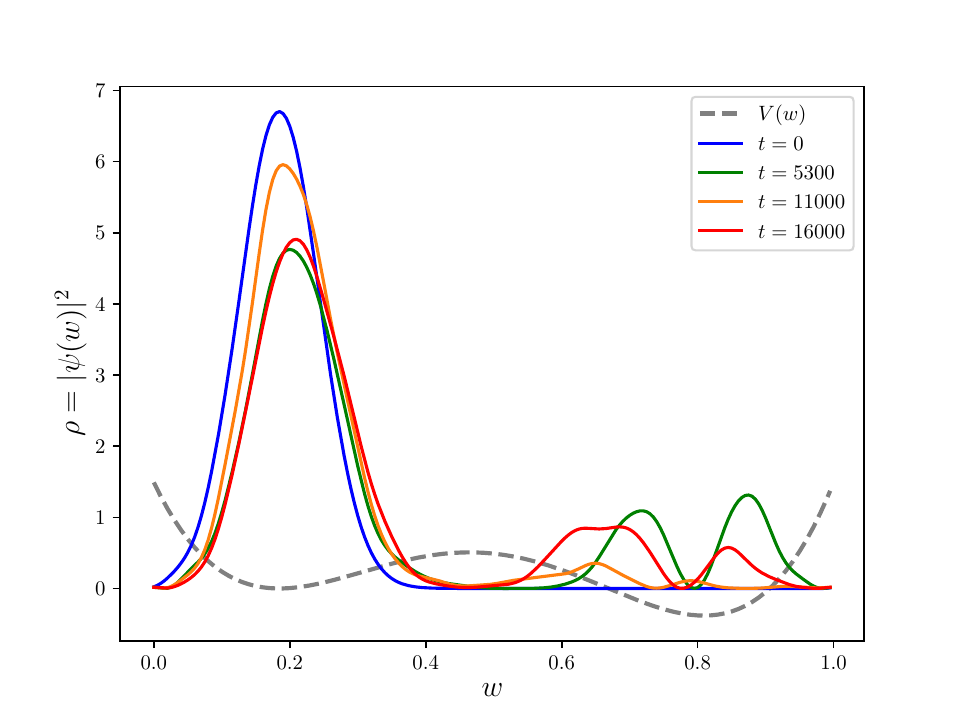}
\includegraphics[width=0.5\textwidth]{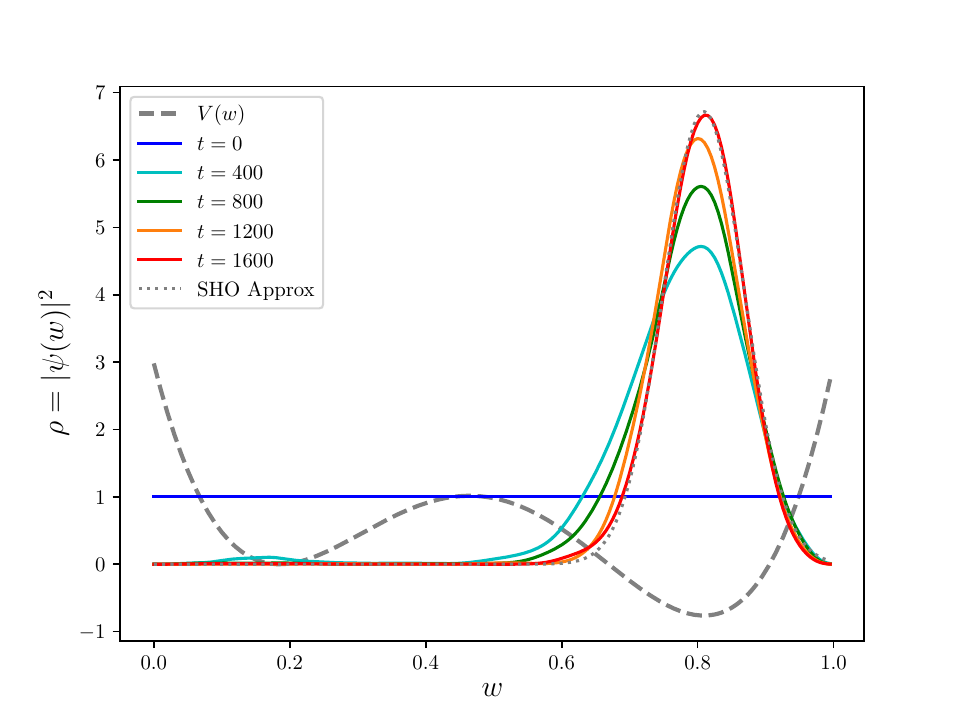}

  \caption{Tunnelling versus adiabatically evolving the ground state in a quartic potential. Here for tunnelling the initial wavefunction is chosen to be the groundstate of the approximate SHO potential around the false minimum (with $\lambda = 4, ~m=100$). For the adiabatic evolution we take $\lambda = 8, m=200$ to ensure a localized peak at the origin. We also show (overlaid dotted line) the groundstate of the SHO approximation obtained by expanding around the global minimum at $w=0.8$. } 
  \label{fig:tunnelling}
\end{figure} 

\begin{figure}
  \centering
\includegraphics[width=0.5\textwidth]{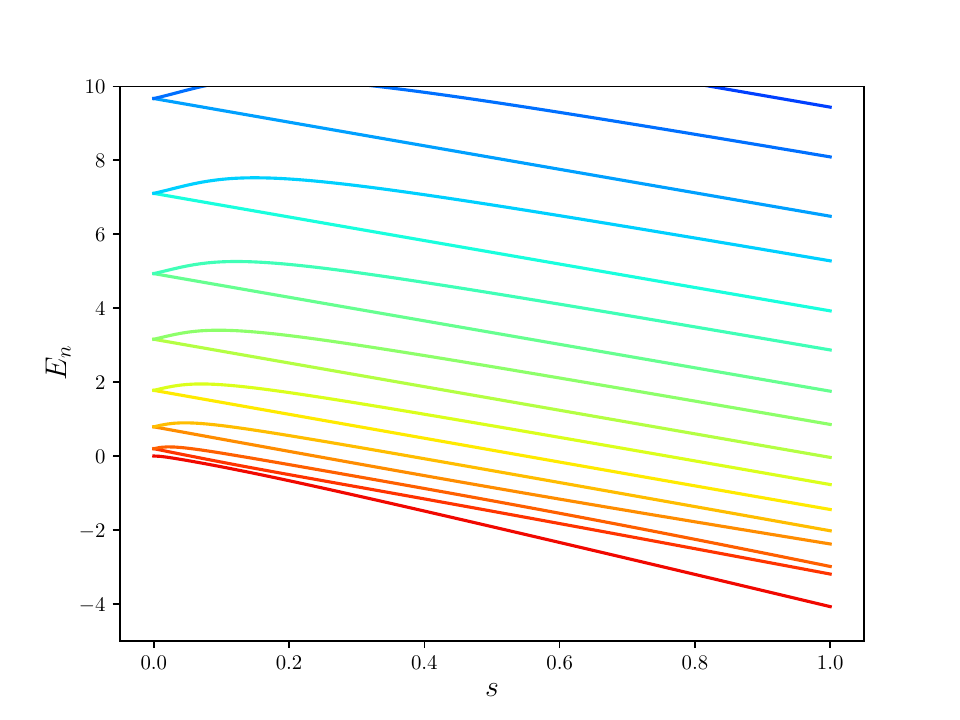}
  \caption{Energies during the adiabatic evolution in Fig.~\ref{fig:tunnelling} showing the isolated ground-state energy. } 
  \label{fig:energies}
\end{figure} 

\subsection{The Pauli-Spin Method}

Despite the straightforward nature of the matrix method for adiabatic evolution,  it is not the most convenient approach for NN training because the matrix $H_{\ell n}$ would grow exponentially with the number of variables (i.e. weights) in the system, due to us being required to store a $2^{N_{\text{}}} \times 2^{N_{\text{}}}$ matrix in general. Instead, it is typically more efficient (we will make a more detailed comparison of the relative efficiencies later in Subsection~\ref{subsec:scaling}) to use the ``Pauli-spin method'': in this method, the variables and hence the Hamiltonian are encoded in a binary fashion in the eigenvalues of Pauli-spins. 

That is, we assign bin values for the variables themselves instead of the wave function by defining the binary $T$ operators 
as in Eq.~\eqref{eq:binary_def}. For example, in the single variable case, we encode 
$w$ discretely as a fractional binary composed of $N$ of the binary spins, $T_\ell$. Hence the operator corresponding to $w$ is 
\begin{equation}
 \hat w ~=~ {2^{-N}} \sum_{\ell = 0}^{N-1} 2^{\ell} T_\ell~.
 \label{eq:weight-encoding}
\end{equation}
 The above encoding yields binned values for possible measurements of the variable, $ \langle \hat w\rangle  \in \{ w_r\} =  \{ 0, \frac{1}{2^{N}},\frac{2}{2^{N}} \ldots , 1- \frac{1}{2^{N}} \}$.
 Thus any particular state $|\psi\rangle $ is defined as 
 \beq
|\psi \rangle ~=~ \sum_r |w_r\rangle \langle w_r | \psi\rangle ~,
 \eeq
with $r= 0\ldots 2^{N-1} $ labelling the possible bin values $w_r$, and with $\rho(w_r) = |\langle w_r | \psi\rangle|^2$ yielding the probability for measuring the state in that particular bin. Essentially this replaces the momentum truncation with a direct variable discretisation. 

This is the general structure for encoding variables. How should we now go about constructing the adiabatic evolution? For the target Hamiltonian $\hat H$, the main aspect to note is that in this discretised variable formulation of the problem, the momentum and hence the kinetic $\hat p^2/2m$ terms would be hard to encode (such terms would have to be encoded by the finite difference which would greatly complicate the Hamiltonian). However, we also note that the kinetic terms in the Hamiltonian did not serve much purpose in locating the global minimum of $V(w)$ anyway: all they do is provide spread in the profile of the eventual ground state. Indeed from Fig.~\ref{fig:groundstate-mass}, it is clear that if we were to take the limit $m\to \infty$ keeping $V(w)$ unchanged, then the final wavefunction would be a spike at the global minimum, which would for optimisation be virtually the ideal outcome. Thus, to determine the global minimum of a potential $V(w)$, we may delete the kinetic terms and set 
\beq 
\hat H ~ = ~ V(\hat w) ~,
\label{eq:H_2}
\eeq
where now the operator $\hat w$ is to be replaced by its encoding in terms of $Z_\ell$ spins given in Eq.~\eqref{eq:weight-encoding}. Note that, unlike the matrix approach, we are now constrained to consider polynomial potentials. Moreover, a modest amount of reduction can be performed on the Hamiltonian. For example, upon expanding the polynomial $\hat V$, we may find powers of Pauli matrices that can be reduced using $T_\ell T_\ell = T_\ell$. (Such reduction is more significant when fewer qubits are used to define each $\hat w$).

To play the role of the trivial Hamiltonian in the adiabatic evolution, $\hat H_0$, we can use the commonly adopted transverse AQC choice,
\beq 
\hat H_0 ~=~ \frac{1}{2} \sum_{\ell=0}^{N-1}\left({\mathbbm 1} - X_\ell \right)~,
\label{eq:H0_2}
\eeq 
where $X_\ell$ is the $X$ Pauli-spin matrix for the $\ell$'th qubit. It is easy to see that $\frac{1}{2^{N/2}}\prod_\ell (|0\rangle_\ell + |1\rangle_\ell)  $ is the groundstate of this Hamiltonian (because $X(|0\rangle + |1\rangle) = (|0\rangle + |1\rangle) $). Expanding, we see that this is the state with degenerate probability in each $w$ bin, which has $\langle \hat H_0 \rangle =0$.

\begin{figure}
  \centering
\includegraphics[width=0.5\textwidth]{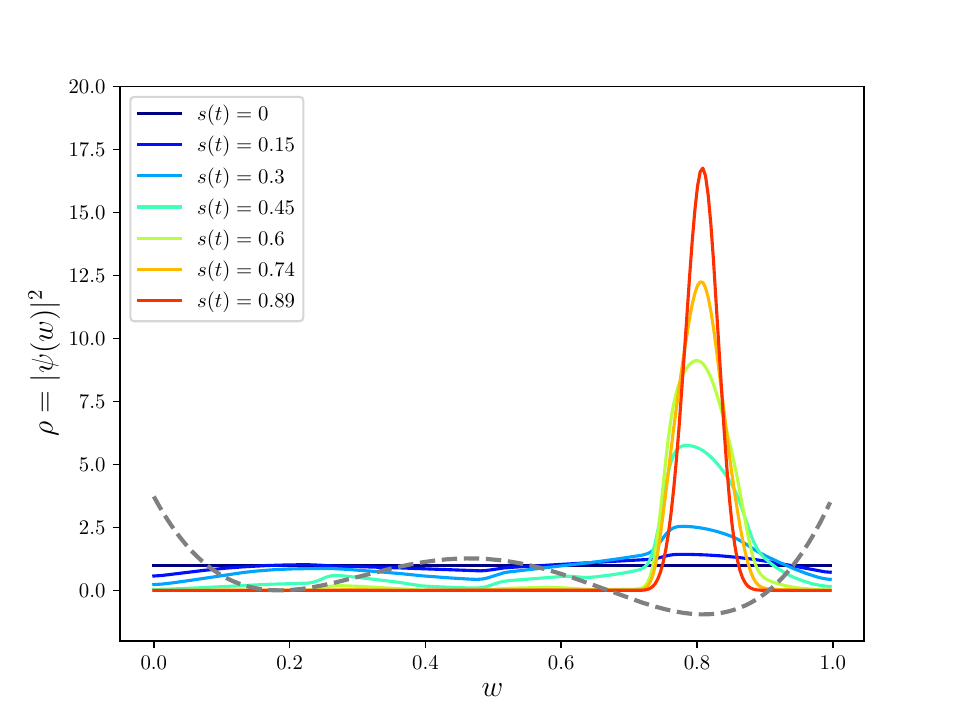}
  \caption{Adiabatically evolving to find the global minimum of the quartic potential using a Pauli-spin encoding of $w$. Here we show the evolving smoothed histogram of the groundstate $\rho(w)\equiv |\psi (w)|^2$ with $w$ encoded in $N = 7$ qubits.  } 
  \label{fig:global-quartic}
\end{figure} 

Finally, we put these two Hamiltonian components, namely $\hat H_0$ of Eq.~\eqref{eq:H0_2} and  $\hat H$ of Eq.~\eqref{eq:H_2}, into the adiabatic evolution equation in Eq.~\eqref{eq:HA_1}, and evolve the system from the initial $\hat H_0$ ground state using the Trotterized circuit generated by {\tt qibo}. The result for the quartic potential is shown in Fig.~\ref{fig:global-quartic}.
As expected, it is highly peaked around the global minimum.

\section{Neural Network training}
\label{NNs}

\begin{figure*}
    \centering
    \includegraphics[width=\textwidth]{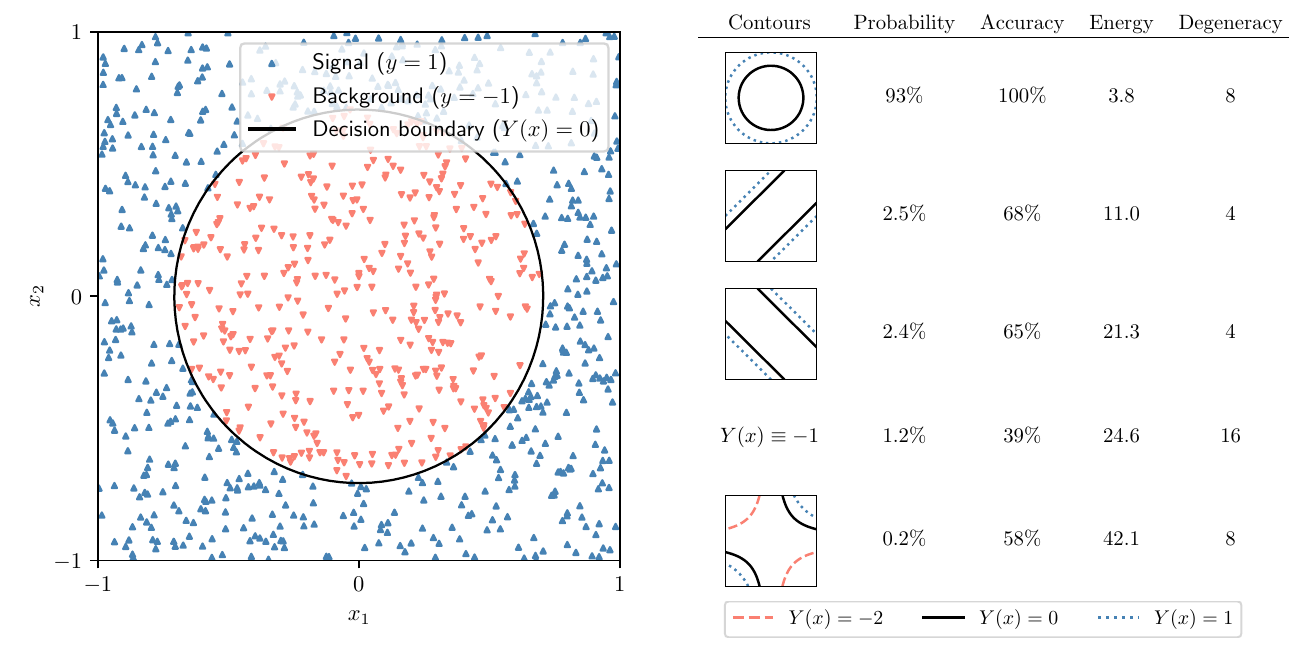}
    \caption{Left: circle dataset and the corresponding decision boundary generated by the most probable final state after adiabatic evolution and measurement. Right: summary of some of the potential outcomes of the final measurement, including the corresponding $Y(x) =$ constant contours, the probability of measuring each of them, their energy, and the degeneracy (the number of equivalent states that generate the same $Y(x)$ function).}
    \label{fig:circle-nn-fig}
\end{figure*}

\subsection{General method}

In this section, we will demonstrate that the AQC optimisation algorithm outlined in the previous section can be used to train machine-learning models, where we will now replace the single $\hat w$ operator with a large number of weights and biases. 
We focus on the supervised learning framework, which aims to find a function $Y(x)$ that approximately reproduces a given set of outputs $y_a$ from a given set of inputs $x_i$.
A classification problem is when the outputs, called labels in that context, take values in a small discrete set.
Otherwise, the problem becomes general non-linear regression.

A machine learning model is a family of functions from which the optimal $Y(x)$ for the available data has to be selected.
The process of finding this optimal function is known as training, and it is typically done by minimising a \emph{loss function} $\mathcal{L}$, which measures the deviation of the predictions $Y(x_a)$ from the labels $y_a$.
For example, one may define it as the mean squared error
\begin{equation}
    \mathcal{L} = \frac{1}{N} \sum_{a = 1}^N \left(Y(x_a) - y_a\right)^2.
\end{equation}
Some of the most versatile models in this setting are neural networks, which are constructed as the composition of layers $L_k(z)$, with each layer given by an affine transformation followed by the element-wise application of a non-linear functions $f_k$:
\begin{align}
    Y(x) &= L_n(\cdots L_1(x)) \\
    L_k(z) &= f_k\left(\sum_j w^{(k)}_{ij} z_j + b^{(a)}_i\right).
\end{align}
The parameters $w$ and $b$ are known as the weights and biases, and the functions $f$ are called the activation functions.

Various classical algorithms have been developed to optimise the loss function $\mathcal{L}$.
Most of them are local optimisation methods, in which the weights and biases are updated iteratively in small increments.
A common problem these algorithms can only partially address is that they can become trapped in local minima for a non-convex loss function.
Thus, quantum algorithms capable of avoiding this problem by directly tunnelling or adiabatically evolving towards the global minimum would work qualitatively differently from classical gradient-based optimisation methods and prevent these problems.

In Section~\ref{AQC}, we outlined two general methods for minimising arbitrary functions, the ``matrix method'' and the ``Pauli-spin method'' .
We shall now apply the Pauli-spin method to minimise the loss as a function of the free parameters of the neural network, which are the weights and biases.

\begin{figure*}
    \centering
    \includegraphics[width=\textwidth]{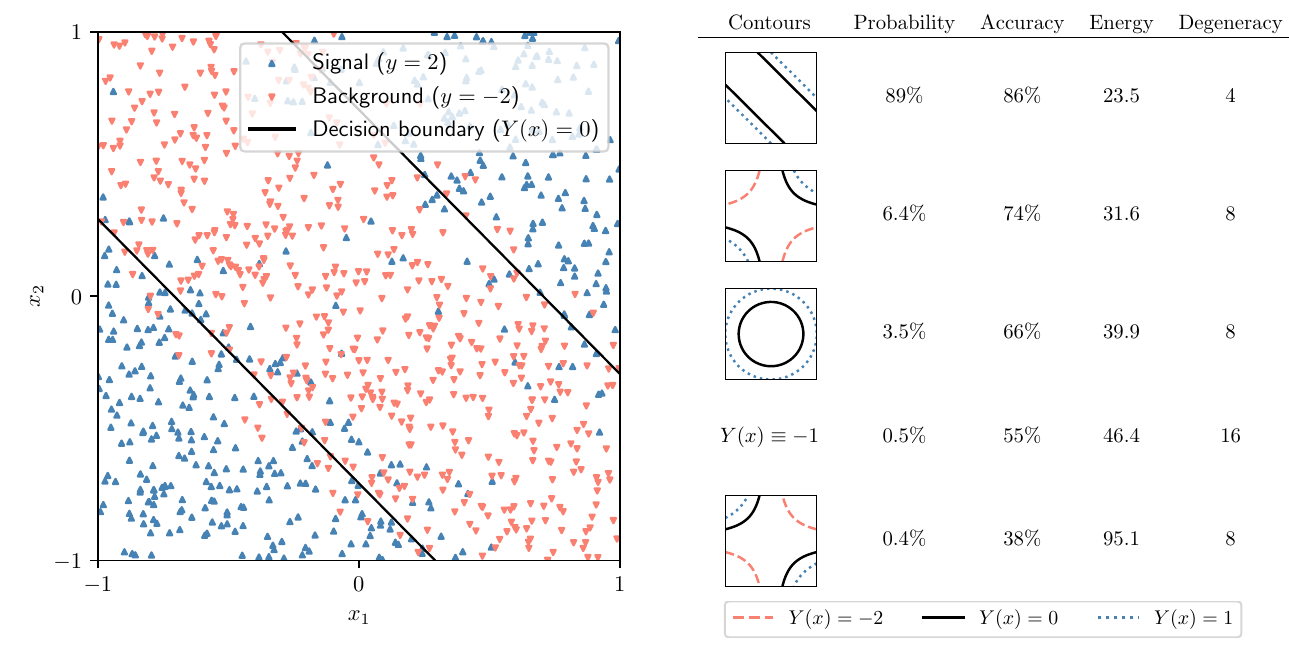}
    \caption{left: band dataset and the corresponding decision boundary generated by the most probable final state after adiabatic evolution and measurement. Right: summary of some of the potential outcomes of the final measurement, including the corresponding $Y(x) =$ constant contours, the probability of measuring each of them, their energy, and the degeneracy (the number of equivalent states that generate the same $Y(x)$ function).}
    \label{fig:band-nn-fig}
\end{figure*}

To begin with, let us make some general remarks on the advantages and disadvantages of the two methods in the neural-network context. 
As we saw, the Pauli-spin method enables an efficient representation of the Hamiltonian in terms of Pauli matrices at the price of approximating the function through polynomials. In the context of neural networks, this implies that the activation function must be approximated by a polynomial, such that the loss function becomes a polynomial in spin matrices of degree given by the number of layers and the degree of the activation function.
One then only needs to store the non-vanishing coefficients of this polynomial.
This can be a significant advantage over matrix encoding.
The effects of the polynomial approximation can be made arbitrarily small
because any well-behaved activation function can be approximated arbitrarily well by a polynomial in a bounded domain. The range of values of the inputs to each activation is bounded and known in advance, given the range of values of the inputs $x$ and the binary-encoded parameters $w$ and $b$. 
The downside of the ``Pauli-spin method'' is that the nested non-polynomial activation functions result in a large gate depth. We will make more quantitative comparisons of the methods later in Subsection~\ref{subsec:scaling}.

\subsection{Toy example}

For concreteness, we will focus on a toy example, although the method can be used to train any other neural network.
Our neural network has two layers, the first mapping 2D points to 2D points and the second mapping 2D points to numbers.
We take activation functions to be $f_1(x) = x^2$ and $f_2(x) = x$, and the biases $b^{(1)}_i = 0$ and $b^{(2)} = -1$.
The output is then given by
\begin{equation}
    Y(x) = \begin{pmatrix}
        w^{(2)}_1 & w^{(2)}_2
    \end{pmatrix} \left[
    \begin{pmatrix}
        w^{(1)}_{11} & w^{(1)}_{12} \\
        w^{(1)}_{21} & w^{(1)}_{22}
    \end{pmatrix}
    \begin{pmatrix}
        x_1 \\ x_2
    \end{pmatrix}
    \right]^2
    - 1,
\end{equation}
where the square is to be understood as the element-wise square function applied to a 2-vector.
We use the Pauli-spin method, with one qubit per parameter only.
This leads to a system with a total of 6 qubits, which allows us to simulate it on a small classical computer using \texttt{qibo} as described in the previous section.

We will use this network to perform a binary classification task, predicting a point to be signal if $Y(x) \geq 0$ and background otherwise.
We therefore call the $Y(x) = 0$ contour the decision boundary.
The simple structure we have chosen allows for several shapes of the decision boundary, from which the optimal one is to be selected by the adiabatic computation.

The two datasets we consider are a set of 1000 randomly chosen 2D points, with uniform distribution in the square $[-1, 1] \times [-1, 1]$.
In the first one, which we call the circle dataset, these 2D points are labelled as $y = 1$ (signal) if $x^2 + y^2 > 1/2$ and $y = -1$ (background) otherwise.
The optimal decision boundary for the circle dataset is thus the circle $x^2 + y^2 = 1/2$, which our toy neural network can achieve.
In the second dataset, which we call the band dataset, they are labelled $y = 2$ (signal) with probability given by $\operatorname{max}[1, (x + y)^2]$ and with $y = -2$ otherwise.
We make this choice so that the data is not perfectly separable, but our neural network can achieve the lowest value of the loss function when it generates a decision boundary of $2 (x^2 + y^2) = 1$

\begin{figure*}
    \centering
    \includegraphics[width=0.9\textwidth]{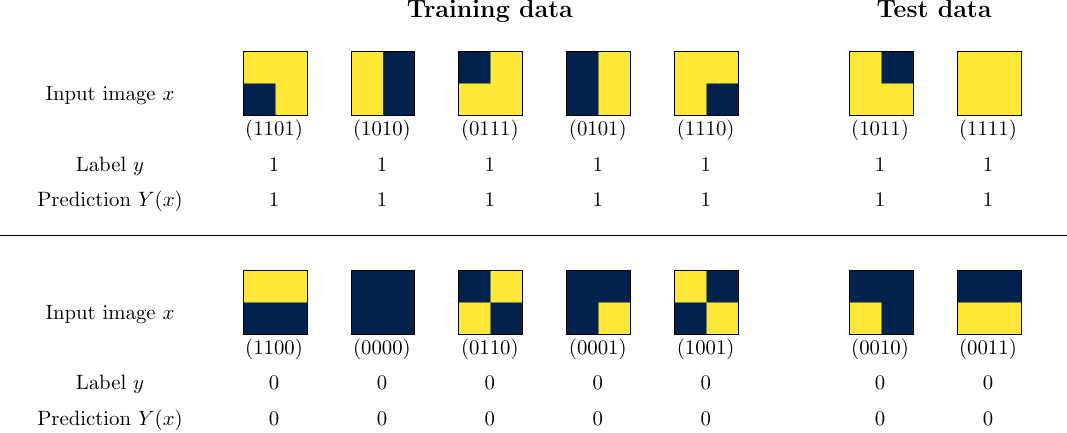}
    \caption{Dataset and predictions from an adiabatically-trained binary neural network. The predictions are generated using the weights determined by the most likely outcome after the final measurement.}
    \label{fig:bnn-fig}
\end{figure*}

To train the neural network, we generate the target Hamiltonian $\hat H$ by replacing each weight in the loss function $\mathcal{L}$ by a $Z$ Pauli matrix.
The initial trivial Hamiltonian $\hat H_0$ is given by Eq.~\eqref{eq:H0_2}.
We use \texttt{qibo} to simulate the adiabatic time evolution in 10 steps from $t = 0$ to $t = 10$, with a linear schedule $s(t) = t/10$.
The final state consists of a superposition of different computational-basis states.
In a real-world device, one would measure all of the $Z_\ell$ to obtain the classical values of the weights in the network.
Our simulation shows that the correct contour for the circle dataset, displayed on the left in Fig.~\ref{fig:circle-nn-fig}, is the most likely outcome of this measurement, with a 93\% probability.
Similarly, the most likely outcome for the band dataset, with 89\% probability, is the optimal contour, shown on the left in Fig.~\ref{fig:band-nn-fig}.
In practice, performing a low number of AQC runs and selecting the final state with the least energy is a viable strategy.

It should be noted that, like most neural networks, the one we are considering has multiple symmetries because different possible values of the weights give rise to the same function $Y(x)$.
An example of such a symmetry consists of flipping both $w^{(1)}_{11} \to -w^{(1)}_{11}$ and $w^{(1)}_{12} \to -w^{(1)}_{12}$.
Two states related to these symmetries must have the same energy under the target Hamiltonian $\hat H$.
On the right side of Figs.~\ref{fig:circle-nn-fig} and~\ref{fig:band-nn-fig}, we have collected the total probability of measuring any of the states leading to each of the most likely $Y(x)$ functions.

One of the consequences of these symmetries is that the minima of the loss function are degenerate, and therefore we are morally in the degenerate minima situation of Section~\ref{AQC}.
In the classical setting, the random initial seed of the optimization algorithm would select one of the degenerate minima.
However, guided by the discussion in section.~\ref{AQC} it is clear that quantum training leads to a different situation, in which the final quantum state is in a superposition of the degenerate minima, all of which have equal probability.
It is thus the final measurement that plays the role of randomly selecting one of the minima. Moreover, it is clear that, generally, one cannot take many measurements and use the expectation values of the weights for the classical values because this would incorrectly average over these degenerate possibilities.

\subsection{Binary neural networks}

\begin{figure*}
    \centering
    \includegraphics[width=0.8\textwidth]{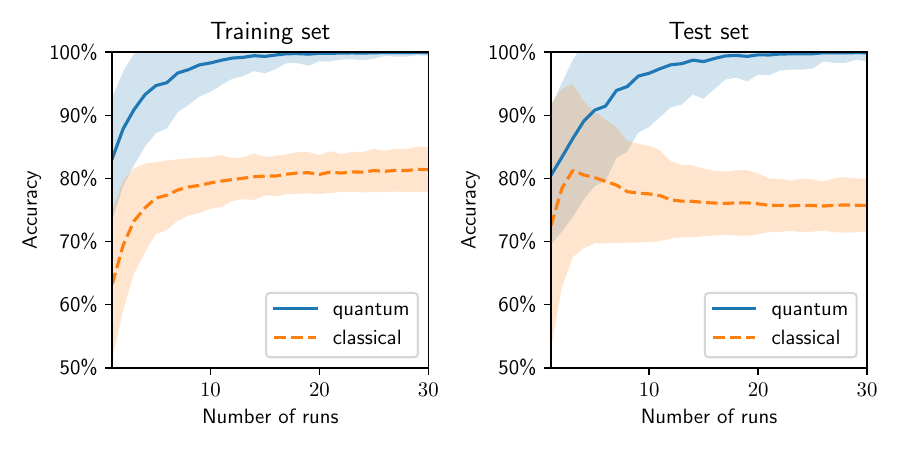}
    \caption{Binary neural network accuracies in the training (left) and test (right) sets from the weights generated by running the training several times and selecting those with the best performance in the training set as a function of the number of runs. The central lines indicate the average accuracy value, and the bands show the 1-standard deviation interval (both computed by repeating the process 1000 times for each value of the number of runs).}
    \label{fig:bnn-accuracies}
\end{figure*}

The limited number of qubits available in current quantum computers makes it more interesting to consider them for training smaller machine-learning models.
A valuable class of such models, with many real-world applications, are binary neural networks~\cite{qin2020binary, yuan2023comprehensive}.
These are neural networks in which the weights can only take the values 0 or 1, the biases are set to 0, and the activation functions are given by
\begin{equation}
    f\left(\sum_{j = 1}^n w_{ij} x_j\right)
    ~=~
    \Theta\left(\sum_{j = 1}^n w_{ij} x_j - \frac{n}{2}\right),
\end{equation}
where the $w_{ij}$ and the $x_j$ are the weights and inputs of the corresponding layer, and $\Theta$ is the Heaviside step function.
The $i$th output of each layer is 1 if at least half of the terms $w_{ij} x_j$ are 1, and zero otherwise.

Binary neural networks can be directly encoded in quantum computers without any loss of expressiveness that we encountered with a polynomial approximation of activation functions and with the discretisation of continuous weights.
The model trained in a quantum device can be {\it exactly the same} as the one implemented in a classical computer.
This can be easily seen by noting that the binary 0/1 weights can be encoded using the binary $T_\ell$ operators constructed from  Pauli $Z_\ell$ matrices via Eq.~\eqref{eq:binary_def}. The activation functions can be viewed as polynomials in the $T_\ell$'s, through the identity
\begin{align}
    f\left( T_i\right)
    =
    \sum_{m = 0}^{\lfloor n / 2 \rfloor} \sum_{i_1 < \ldots < i_m}
        \prod_{j \neq i_1, \ldots, i_m} T_j
        \prod_{k = i_1, \ldots, i_m} \bar{T}_k,
    \label{eq:poly-theta}
\end{align}
where $\bar{T} = 1 - T$.
For example
\begin{align}
    f(T_1, T_2, T_3)
   &~ = ~
    T_1 T_2 T_3 + T_1 T_2 \bar{T}_3 \nonumber \\
    &~\qquad + T_1 \bar{T}_2 T_3 + \bar{T}_1 T_2 T_3~.
\end{align}

The discrete nature of binary neural networks makes them even more difficult to train with conventional classical methods, which are, as we have seen, typically based on gradient descent.
Adiabatic quantum training completely avoids this issue, as it can be done using the same procedure we outlined above for quasi-continuous neural networks.

Since the outputs are binary (and assuming that the labels $y$ are binary as well), one can use a simpler linear loss function,
\begin{equation}
    \mathcal{L} ~=~ \sum_a (-1)^{y_a} Y(x_a)~.
    \label{eq:linear-loss}
\end{equation}
With such a loss function those points  $x_a$ with either label, $y_a = 0$ or $y_a = 1$, are penalised by one unit in the loss function if there is an incorrect prediction, $Y(x_a) \neq y_a$. (That is $\{y_a,Y\}=\{0,1\}$ is incorrect and contributes $\Delta {\cal L} = 1$ 
versus $\{y_a,Y\}=\{0,0\}$ which contributes $\Delta {\cal L} = 0$. Likewise $\{y_a,Y\}=\{1,1\}$ is correct and contributes $\Delta {\cal L} = -1$ versus $\{y_a,Y\}=\{1,0\}$ which contributes $\Delta {\cal L} = 0$.)  

To test this approach, we prepare a dataset of images with $2 \times 2$ binary pixels, labelling them with $y = 1$ (signal) if there are two pixels set to 1, one directly above the other, and $y = 0$ (background) otherwise.
We select seven signal and seven background samples to balance the dataset.
We then split the dataset into 5 (signal) + 5 (background) training images to be included in the loss function and 2 + 2 test images to check the generalisation properties of the trained model.
The selection and splitting are done randomly from the 16 possible binary images.
The resulting train/test datasets are displayed in Fig.~\ref{fig:bnn-fig}.

For the binary neural network, we choose one with two layers, with the first having four inputs and two outputs and the second having two inputs and one output.
The total number of weights, which are in one-to-one correspondence with the qubits, is 10.

To train the network we use \texttt{qibo} to simulate an adiabatic computation as in the previous section, with $\hat H$ now determined by substituting into the loss function $\mathcal{L}$ the expression for the weights in terms of qubits, and the polynomial representation of the step function in Eq.~\eqref{eq:poly-theta}.
The predictions generated by the most likely weights after the final measurement are shown in Fig.~\ref{fig:bnn-fig}.
They are 100\% accurate in both the training and the test sets.
The probability of obtaining these perfectly accurate weights in the final measurement is 18\%.
To assess the efficiency of the training this can be compared with the portion of the space of weights that generates such predictions, which is 0.2\%.

Since the best values of the weights are obtained with the highest probability but not with certainty, it is profitable to perform several runs of the adiabatic computation and select the one that results in the highest accuracy in the training set.
In Fig.~\ref{fig:bnn-accuracies}, we show how the accuracy of the trained network on both the training and the test sets improves with the number of runs.
To obtain it, we generate a pool of 1000 sets of trained weights, with distribution given by the final state of the adiabatic evolution, before the final measurement.
For each value of the number of runs $n$ shown in the Fig.~\ref{fig:bnn-accuracies}, we select $n$ sets of weights from the pool and pick the maximum accuracy.
This process is repeated 1000 times, and the average and standard deviation of the resulting accuracies are displayed in the figure.

We compare it with the accuracy of a classical analogue trained using the Adam gradient descent algorithm.
To construct this analogue, we replace the step functions with sigmoids, replace the binary weights with continuous ones, and add a penalty term to the loss function of the form $w^2 (w - 1)^2$ for every weight.
The effect of this penalty term is to drive the weights to 0 or 1 values.
The classical values displayed in Fig.~\ref{fig:bnn-accuracies} correspond to the same process as for the quantum ones described above, using a pool of 1000 values generated through 1000 classical training runs.

The quantum training exhibits a better performance and generalisation, with the accuracy in both the training and the test sets quickly approaching 100\%; while the classical training tends to get stuck in local minima that lead to accuracies of around 80\% in the training set, with lower ones in the test set, indicating poor generalisation.

\subsection{Comparative estimates of scaling}

\label{subsec:scaling} 

The different approaches to encoding the loss function presented here incur different computational costs in calculating the target Hamiltonian $\hat H$ and other gate complexities in the quantum circuit that implements the adiabatic time evolution.

It is worth comparing the different approaches to see how they scale with meta-parameters, e.g. number of hidden layers, total number of qubits and so forth. 
To do this we assume that $\hat H$ is decomposed as a polynomial in Pauli matrices to encode in a time-evolution circuit.
The number of gates in the circuit will then be bounded from above by a quantity proportional to the number of terms $T$ in this polynomial, multiplied by its degree $D$.

A Hamiltonian for a system $N_q$ qubits is in general an $2^{N_q} \times 2^{N_q}$ matrix.
Thus, for the generic ``matrix approach'', one needs to compute $2^{2 N_q}$ quantities in the preparation stage of the calculation.
The decomposition of $\hat H$ in terms of Pauli matrices will thus require $2^{2 N_q}$ matrix multiplication and trace operations.
Finally, the resulting polynomial in Pauli matrices will have degree $D = 2^{2 N_q}$ and roughly $T = 2^{2 N_q}$ terms, so the number of gates scales roughly as $2^{4 N_q}$.
However, the loss functions of neural networks typically lead to a very sparse $\hat H$, so there is room for significant improvement on these scalings.

Using the ``Pauli-spin method'' is one possible strategy to take advantage of sparsity.
The maximum number of terms in $\hat H$ is then several chains of length $2^{N_q}$ of identity or Pauli $Z$ matrices.
One needs to compute the coefficient to each of these chains in $\hat H$, so the maximum number of quantities to compute in this approach is a factor $2^{N_q}$ smaller than in the general case.
In practice, this number might be much smaller.
Moreover, these quantities are computed directly by replacing the binary encoding of the weights with the loss function, with no need for decomposition of $\hat H$ into a basis involving matrix multiplications and traces.
The degree of the $\hat H$ polynomial is, in this case, independent of the number of qubits and increases with the number of layers of the network, but not with the number of weights per layer.

Thus the scaling of both $T$ and $D$ improves significantly for relatively shallow networks in the Pauli-spin approach.
To simplify the discussion, we consider a neural network with $L$ layers, all having a polynomial activation with degree $d$, and an $M \times M$ matrix of weights with no biases.
The number of terms in the $\hat H$ polynomial is then
\begin{equation}
    T \lesssim M^{d^L}~.
    \label{eq:T-bound}
\end{equation}
This can be shown by induction on $L$.
For a network with a single layer, $L = 1$, the number of terms is bounded by the number of terms in a degree-$d$ polynomial in $M$ variables:
\begin{equation}
    T < \begin{pmatrix} d + M \\ d \end{pmatrix}
    \overset{M \to \infty} \sim M^d~.
\end{equation}
Similarly, adding a layer to an $(L-1)$-layer network gives several terms bounded by the number of terms in a degree-$d$ polynomial in variables that are degree-$M^{d^{L-1}}$ polynomials themselves This equation shows that the number of terms, and thus the gate complexity, is polynomial in $M$ in this approach.
On the other hand, the scaling with the number of layers is much worse: it is doubly exponential.
It will thus quickly saturate the generic bound for Pauli-spin encoded functions of $2^{N_q}$, so the latter is the stronger one for deep neural networks.

In the case of binary neural networks with step-function activations, the degree of the activation polynomials is $d = M$.
Thus, the advantages over classical algorithms provided by their quantum training are obtained at the price of an $M^M$ scaling of the number of terms with the number of weights per layer.
A potential source for improvement on this front is binary activations with a lower degree or a lower number of terms.
An example of such an activation would be one that required all inputs of the layer to be 1 for it to be 1, otherwise being 0.


\section{Conclusions}
\label{conclusions}

Neural networks are ubiquitous optimisation tools in science and everyday tasks. The most time and resource-consuming part of their design is the training process. In this study, we have demonstrated the potential of Adiabatic Quantum Computing as a powerful tool for training neural networks. Our work addresses the computational challenges encountered when classically training NNs. We have demonstrated that AQC can effectively be implemented on gate quantum computers to train neural networks with continuous and discrete weights, as well as so-called binary networks.
Our findings indicate that AQC offers a robust and efficient approach to finding the global minimum of the loss function, thereby optimising the NN. It is then possible to extract the optimally trained network parameters for deployment as a classical neural network.

The proposed methodology involving the "matrix method" and the "Pauli-spin method" effectively encodes and solves this optimisation problem. As we leveraged the {\tt qibo} package to facilitate fast and accurate quantum circuit evaluation, our approach is scalable and practical for near-term quantum devices. 

Compared to previous quantum approaches, which were based on quantum annealing using a transverse Ising model Hamiltonian, the AQC approach that we have proposed in this paper enhances the expressivity of the trained neural networks and expands the applicability of quantum training methods to the gate quantum computing paradigm.

Extending this methodology to more complex neural network architectures and loss functions would be of interest in expanding its applicability to broader classes of problems. Thus, this approach opens up new avenues for harnessing the computational prowess of quantum computation in the realm of machine learning, particularly in the training of neural networks. 

\vspace{3mm}
\noindent {\bf {Acknowledgements}:}  We would like to thank Luca Nutricati for helpful discussions and Stefano Carrazza and Matteo Robbiati for help with {\tt qibo}. S.A. and M.S. are supported by the STFC under grant ST/P001246/1. J.C.C. is supported by the Spanish Ministry of Science and Innovation, under the Ram\'on y Cajal program.

\bibliography{}
\end{document}